\documentclass[twocolumn, prl, letterpaper, superscriptaddress, showpacs]{revtex4}
\usepackage{amssymb, amsmath,amsfonts}
\usepackage[dvips]{graphicx}
\usepackage{color}

\newcommand\degrees[1]{{#1}$^{\circ}$}
\newcommand\etal{\textit{et~al.}}
\newcommand\celsius[1]{\degrees{#1}C}
\newcommand\kelvin[1]{\mut{#1}{K}}
\newcommand\mJcm[1]{\mut{#1}{mJ~cm^{-2}}}
\newcommand\ohmcm[1]{\mut{#1}{\,\Omega^{-1}\,cm^{-1}}}
\newcommand\nm[1]{\mut{#1}{nm}}

\newcommand\fs[1]{\mut{#1}{fs}}

\newcommand\mathenv[1]{$#1$}
\newcommand\chem[1]{$\mathrm{#1}$}
\newcommand\mut[2]{\mathenv{#1\,\,\mathrm{#2}}}
\newcommand\figref[1]{Figure~\ref{#1}}

\newcommand\percent[1]{#1\%}

\begin{document}
\title{Softening of the insulating phase near $T_c$ for the photo-induced insulator-to-metal phase transition in
vanadium dioxide}
\author{D.~J.~Hilton}
\altaffiliation{Present Address: Department of Physics, University
of Alabama-Birmingham, 1530 3rd Ave S, Campbell Hall 310,
Birmingham, AL 35294-1170}
\affiliation{Center for Integrated
Nanotechnologies, MS K771, Los Alamos National Laboratory, Los
Alamos, NM 87545, USA}
\author{R.~P.~Prasankumar}
\affiliation{Center for Integrated Nanotechnologies, MS K771, Los
Alamos National Laboratory, Los Alamos, NM 87545, USA}
\author{S.~Fourmaux}
\affiliation{Universit\'{e} du Qu\'{e}bec, INRS-\'{E}nergie et
Mat\'{e}riaux et T\'{e}l\'{e}communications, Varennes, Qu\'{e}bec,
Canada, J3X 1S2}
\author{A.~Cavalleri}
\affiliation{Department of Physics, Clarendon Laboratory, University
of Oxford, Parks Rd. Oxford, OX1 3PU, United Kingdom}
\author{D.~Brassard}
\affiliation{Universit\'{e} du Qu\'{e}bec, INRS-\'{E}nergie et
Mat\'{e}riaux et T\'{e}l\'{e}communications, Varennes, Qu\'{e}bec,
Canada, J3X 1S2}
\author{M.~A.~El Khakani}
\affiliation{Universit\'{e} du Qu\'{e}bec, INRS-\'{E}nergie et
Mat\'{e}riaux et T\'{e}l\'{e}communications, Varennes, Qu\'{e}bec,
Canada, J3X 1S2}
\author{J.~C.~Kieffer}
\address{Universit\'{e} du Qu\'{e}bec, INRS-\'{E}nergie et Mat\'{e}riaux et T\'{e}l\'{e}communications, Varennes, Qu\'{e}bec, Canada, J3X
1S2}
\author{A.~J.~Taylor}
\affiliation{Center for Integrated Nanotechnologies, MS K771, Los Alamos National Laboratory, Los Alamos, NM 87545, USA}
\author{R.~D.~Averitt}
\email{raveritt@physics.bu.edu}
\altaffiliation{Present Address: Department of Physics,
Boston University, 590 Commomwealth Ave., Boston, MA 02215}
\affiliation{Center for Integrated Nanotechnologies, MS K771,
Los Alamos National Laboratory, Los Alamos, NM 87545, USA}
\date{\today}

\begin{abstract}
We use optical-pump terahertz-probe spectroscopy to investigate the
near-threshold behavior of the photoinduced insulator-to-metal (IM)
transition in vanadium dioxide thin films. Upon approaching $T_c$ a
reduction in the fluence required to drive the IM transition is
observed, consistent with a softening of the insulating state due to
an increasing metallic volume fraction (below the percolation
limit). This phase coexistence facilitates the growth of a
homogeneous metallic conducting phase following superheating via
photoexcitation. A simple dynamic model using Bruggeman effective
medium theory describes the observed initial condition sensitivity.
\end{abstract}

%(72.80.Ga) Transition-metal compounds  (78.20.-e) Optical constants
%(including refractive index, complex dielectric constant,
%absorption, reflection and transmission coefficient, emissivity)
%
%\pacs{(71.30.+h) insulator-to-metal transitions and other electronic
%transitions  (78.47.+p) Time-resolved optical spectroscopies and
%other ultrafast optical measurements in condensed matter}
\date{\today}
\maketitle

Strongly correlated electron materials are characterized by extreme
sensitivity to external stimuli which result from the subtle
interplay between many degrees of freedom with comparable energy
scales (lattice, orbital, electronic and magnetic). High temperature
superconductivity, colossal magnetoresistance, ferromagnetism, and
insulator-to-metal transitions are some of the best known examples
of the exotic behavior exhibited by these materials.

During the past several years, time-resolved optical studies have
been utilized to investigate dynamics associated with the competing
degrees of freedom in these materials \cite{AverittCMRReview}. In
many of these studies the goal is to interrogate the dynamics within
a particular phase.  However, photo\-induced phase transitions
provide an important complementary approach to investigate the
physical pathway connecting different correlated electron states as
well as their mutual competition \cite{PIPTspecialissue}.

A photo\-induced phase transition can arise following impulsive
heating or photo-doping and provides a means of controlling the
overall phase of a solid on an ultrafast timescale
\cite{CavalleriVO2,colletScience,iwaiPRL}. One model system used to
study insulator to metal transitions in correlated electron systems
is vanadium dioxide ($\mathrm{VO_2}$). This material undergoes an
insulator to metal transition when heated above \kelvin{340}
accompanied by a structural distortion from a monoclinic to a rutile
phase\cite{VO2GoodenoughStructuralPhaseTransition,allenPRL,zylb_mott_PRB1975}.
Time-resolved optical studies can provide insight into its origin
and its technological
potential\cite{balbergCLMM,roachAPL1971,riniOL2005}.

The interplay between band and Mott-insulating behavior in
$\mathrm{VO_2}$ has been addressed in the time domain, where a
limiting structural timescale for the creation of the metallic phase
suggests the importance of band insulating character for
$\mathrm{VO_2}$ \cite{CavalleriVO2StructuralPhaseTransition}.
Advances in ultrafast technology have also enabled more
comprehensive investigations involving, most recently, femtosecond
x-ray absorption spectroscopy
\cite{cavalleriPRB2004,cavalleriPRL2005}.
%%%
%%%However, one direct measurement of the ultrafast insulator-metal
%%%transition, the conductivity response at THz frequencies, has never
%%%been performed.

In this Letter, we measure the time-dependent conductivity of
$\mathrm{VO_2}$ during a photo\-induced insulator-metal transition.
Our dynamic experiments focus on the poorly understood
near-threshold behavior, where phase separation and domain formation
inhibit the measurement of conventional transport properties since a
macroscopic conductivity pathway is not established
\cite{fiebig1998Science}. As a function of increasing initial
temperature (up to $\sim T_c$) we observe a reduction in the
deposited energy required to drive the IM transition. Such a
response indicates an initial condition sensitivity which we
interpret as a softening of the insulating state due to the
existence of metallic precursors. These precursors facilitate the
growth of a homogeneous metallic conducting phase following
superheating. A simple dynamic model using Bruggeman effective
medium theory describes the observed response. We emphasize that
this percolation model is valid near threshold whereas, at higher
fluences, the transition is prompt and nonthermal
\cite{CavalleriVO2StructuralPhaseTransition,cavalleriPRB2004,cavalleriPRL2005}. Our results may be relevant in other systems which exhibit tendencies towards phase separation such as spin-crossover complexes \cite{liuSC2003}.

\begin{figure}[b]
\begin{center}
        \includegraphics[scale=0.30]{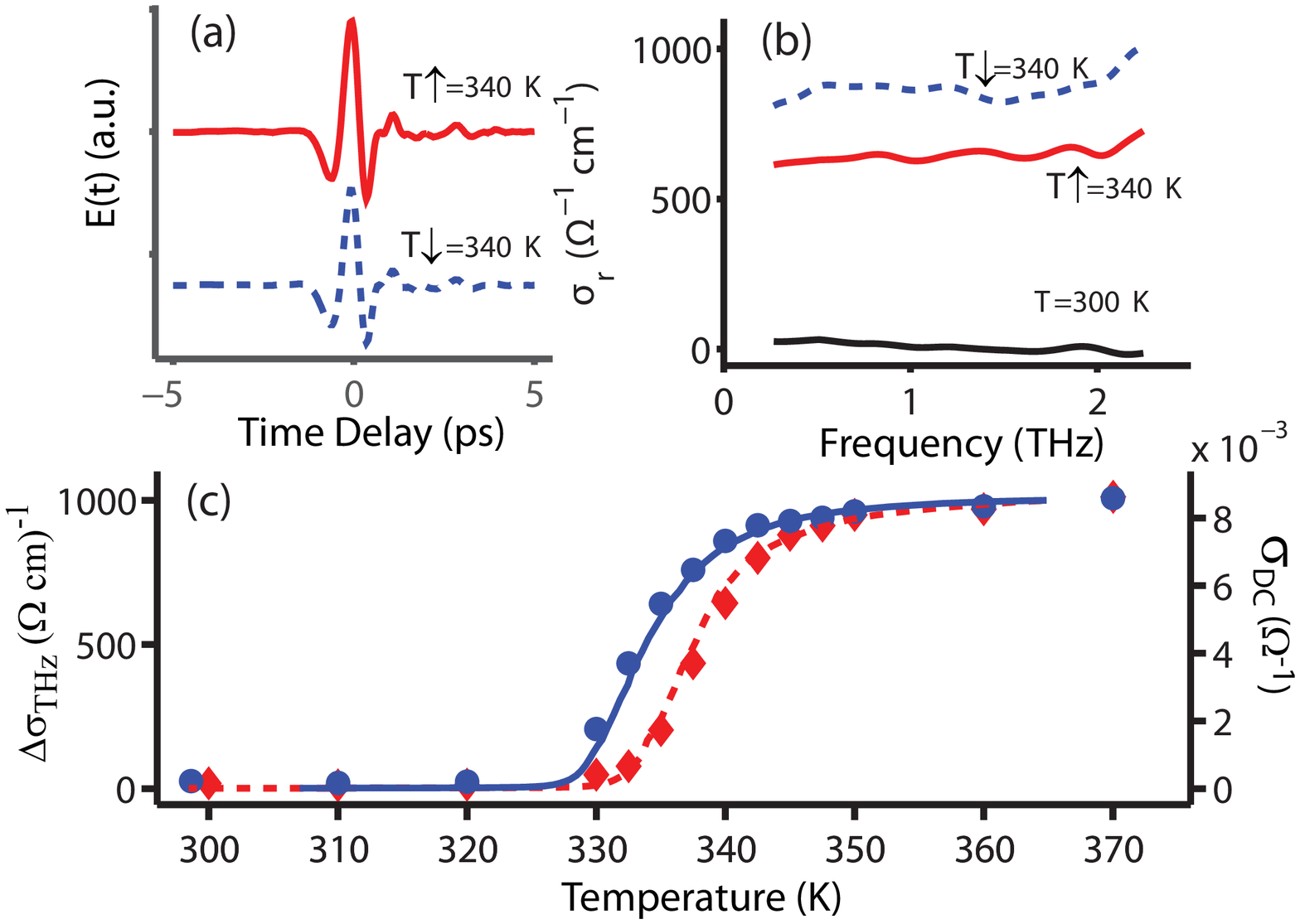}
        \caption{(color online) \label{fig:Sigma} (a) THz waveform transmitted
        through the sample when heated to \kelvin{340} from
        \kelvin{300} (\mathenv{T_\uparrow}) and when cooled from \kelvin{370}
        (\mathenv{T_\downarrow}), offset for clarity (b) The optical
        conductivity vs frequency as determined from the transmitted waveforms in (a) and at
        \kelvin{300}. (c) The temperature-dependent
        hysteresis of the THz conductivity (left scale,
        \mathenv{\blacklozenge\mathrm{~for~}T_{\uparrow}},
        \mathenv{\bullet\mathrm{~for~}T_{\downarrow}}) and of the
        conductance measured using standard DC electrical techniques
        (right scale, \mathenv{\mathrm{--~for~}T_{\uparrow}},
        \mathenv{\mathrm{-~for~}T_{\downarrow}}).
        }
\end{center}
\end{figure}

Our (011) oriented \chem{VO_2} thin film was grown on a (100)
\chem{MgO} substrate by radio-frequency (\mut{13.56}{MHz}) magnetron
sputtering of a vanadium target (\percent{99.99} purity) in a
\chem{Ar} and \chem{O_2} mixture at a pressure of 2 mTorr (chamber
base pressure of \mut{2\times10^{-8}}{Torr}) and a temperature of
\celsius{500} \cite{BrassardVO2Growth}.  The stability of the
deposition process and the thickness of the films (\mut{100}{nm})
were monitored in-situ by a calibrated microbalance and
independently verified using standard ellipsometric techniques.  The
crystalline quality, grain size, and orientation of the \chem{VO_2}
films were characterized by x-ray diffraction.

First, we determined the time-independent THz conductivity of our
$\mathrm{VO_2}$ films as a function of sample temperature. We use
the output of a \fs{50} titanium:sapphire amplifier to generate
nearly-single-cycle THz pulses via optical rectification in ZnTe and
employ standard terahertz time-domain spectrometry to measure the
THz frequency conductivity; further details can be found in
refs.~\onlinecite{SensingWithTerahertzRadiation} and
\onlinecite{THzConductivityAnalysis}. In \figref{fig:Sigma}(a), we
show the transmitted THz waveform at \kelvin{340}, where the top
waveform (displaced for clarity) results when the sample is heated
from \kelvin{300} to \kelvin{340} and the bottom results when the
sample is cooled from \kelvin{370} to \kelvin{340}. From the time
domain data, the calculated conductivity of the $\mathrm{VO_2}$ film
from \mut{0.25}{THz} to \mut{2.25}{THz} is determined
(\figref{fig:Sigma}(b)).
%%% The real conductivity of the film at
%%%\kelvin{340} is frequency independent and approximately \ohmcm{820}
%%%upon cooling from \kelvin{370} while it is \ohmcm{600} when heated
%%%from from \kelvin{300}.

\figref{fig:Sigma}(c) shows the magnitude of the real conductivity
as a function of temperature. We also plot the temperature-dependent
DC electrical conductance measured (on the same film) between two
metal contacts, which shows good agreement with the temperature
dependence of the THz conductivity. The hysteresis is consistent
with the first order phase transition in $\mathrm{VO_2}$ and the
narrow width ($<5~\mathrm{K}$) and magnitude of the metallic state
conductivity attest to the quality of our films \cite{
VO2GoodenoughStructuralPhaseTransition, OriginalVO2MITPaper}.

In order to study the conductivity dynamics of the photo\-induced
phase transition, we excite the sample using 1.55 eV pulses from the
amplified laser and monitor the change in the THz transmission as a
function of the relative delay between the pump pulse and the THz
probe pulse \cite{AverittCMRReview}.  This film is approximately one
optical absorption length thick at \nm{800}, which results in a
nonuniform excitation profile along the direction of propagation;
however, this does not significantly influence our results. We have
verified that the induced change in conductivity is frequency
independent across the bandwidth of our THz pulse indicating uniform
excitation across the THz probe beam.

\begin{figure}[tb]
\begin{center}
        \includegraphics[scale=0.40,angle=0]{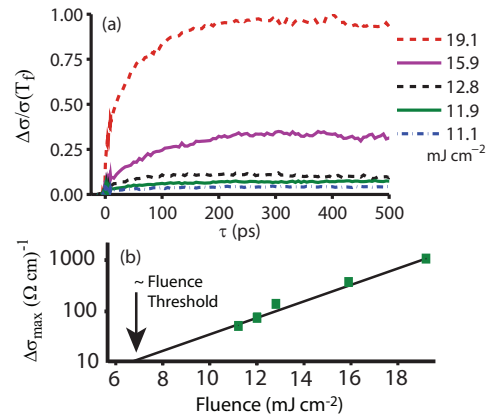}
        \caption{(color online) \label{fig:Fluence} (a) Photoinduced conductivity change at
        \kelvin{300} for various fluences where \mathenv{\sigma\bigl(T_f\bigr)} is the conductivity in
        the full metallic state (b) Magnitude of the conductivity
        change as a function of fluence at \kelvin{300}.}
\end{center}
\end{figure}

\figref{fig:Fluence}(a) shows the time-resolved conductivity as a
function of pump fluence at \kelvin{300}. There is a rise time of
$\sim$100\,ps to obtain the conductivity of the product metallic
phase at all fluences. This is significantly longer than the initial
excitation pulse, meaning that photoconductivity arising from
carriers excited to the conduction band (e.g as occurs in GaAs) is
not responsible for the induced response. Additionally, lattice
heating via carrier thermalization occurs in approximately one
picosecond indicating that the conductivity dynamics are more
complex than simple heating above $T_c$. Crucially, however, for
fluences greater than $\sim$10 mJ cm$^{-2}$, the deposited energy
density ($\sim$ 500 J cm$^{-3}$) is considerably above what is
required to heat above $T_c$ ($\sim$ 200 J cm$^{-3}$ at
\kelvin{300}).

\figref{fig:Fluence}(b) shows the maximum induced conductivity,
\mathenv{\Delta \sigma_{max}}, at each pump fluence extracted from
the data in \figref{fig:Fluence}(a). There is a marked decrease in
the maximum obtainable conductivity with decreasing fluence. At a
fluence of \mJcm{19.2}, the induced conductivity at long times is
\percent{100} of the metallic phase conductivity (i.e. at
\kelvin{370}). Extrapolation of the photo\-induced conductivity
change at \kelvin{300}, as shown in Fig. 2(b), yields a non-zero
fluence threshold of $\sim$\mJcm{7}. The existence of a fluence
threshold is a well-known feature in photo\-induced phase
transitions, where the cooperative nature of the dynamics results in
a strongly non-linear conversion efficiency as a function of the
number of absorbed photons.  In the present case, photoexcitation
leads to a rapid increase of the lattice temperature, initiating the
nucleation and growth of metallic domains which coalesce (i.e.
percolate) to yield a macroscopic conductivity response.
%%%This will be described in more detail below using Bruggeman
%%%effective medium theory.

\begin{figure}[t]
\begin{center}
        \includegraphics[scale=0.35,angle=0]{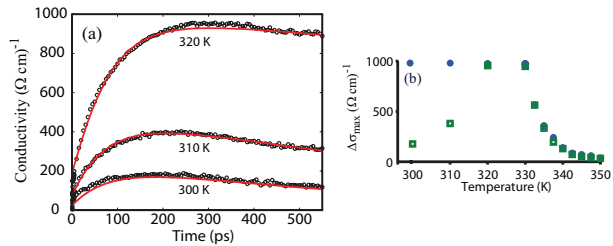}
        \caption{(color online) \label{fig:PumpProbe} (a) Induced conductivity change as a function of time at a fluence
        of \mJcm{12.8} for various initial temperatures. (b) Magnitude of the induced
        conductivity ($\blacksquare$) and the maximum possible conductivity change
        ($\bullet$). The black lines are a fit as described in the text.
        (c) Fluence threshold
        as a function of base temperature.}
\end{center}
\end{figure}

We have also measured the photo\-induced terahertz conductivity as a
function of base temperature. An optical pump fluence of \mJcm{12.8}
was used, less than is required to drive the full metallic
transition at room temperature (though, as described above, more
than enough to superheat the film above $T_c$). The conductivity
dynamics as a function of time at initial temperatures of 300, 310,
and 320~K are displayed in \figref{fig:PumpProbe}(a). From
\figref{fig:PumpProbe}(a) and (b) it is evident that, below $T_c$,
the photoinduced change in the conductivity is less than the maximum
possible induced change. However, with increasing initial
temperature the induced conductivity change increases and obtains
the maximum possible value at \kelvin{320}. Thus, in the insulating
phase, there is a decrease in the threshold to drive the sample
metallic with increasing temperature. At temperatures greater than
\kelvin{330}, the photoinduced change in conductivity follows the
maximum possible induced change (\figref{fig:PumpProbe}(b)). This
occurs since the incident fluence of \mJcm{12.8} is sufficient to
drive the conductivity to its full metallic state consistent with a
decreasing threshold. Finally, in \figref{fig:PumpProbe}(c) we plot
the fluence threshold as a function of base temperature as
determined from several series of data such as that in
\figref{fig:Fluence}(b). This further emphasizes the softening that
occurs in the insulating state with increasing base temperature.

Summarizing the dynamics in \figref{fig:Fluence} and
\figref{fig:PumpProbe}, (i) the conductivity rise time of $\sim$100
ps is substantially longer than the excitation pulse or electron
thermalization time, (ii) a fluence of \mJcm{12.8} heats the sample
well in excess of $T_c$, (iii) despite (ii), the maximum possible
conductivity is not obtained at \kelvin{300} indicating a stiffness
with respect to driving the IM transition and, (iv) this stiffness
towards fully driving the IM transition decreases with increasing
temperature indicating a softening of the insulating phase. While a
complete description of the dynamics is difficult, in the following
we present a simple dynamic model using Bruggeman effective medium
theory that, to first order, describes the experimentally measured
dynamics.

Bruggeman effective medium theory (BEMT)
\cite{BruggemanEffectiveMediumTheory,
StroudBruggemanEffectiveMediumTheory} is a mean field description to
describe inhomogeneous media. For VO$_{2}$, this corresponds to the
coexistence of a metallic volume fraction (\mathenv{f_m}) and an
insulating volume fraction (1-\mathenv{f_m}) which depend on
temperature. Previous work has described the temperature-dependent
conductivity (both the finite transition temperature width and
hysteresis) in terms of the coexistence of metallic and insulating
phases in \chem{VO_2} \cite{VO2GoodenoughStructuralPhaseTransition}.
More recently, additional experimental support for this idea from
time-integrated optical conductivity and scanning probe measurements
has been presented\cite{ChoiMidIRVO2,changeTSF2005}.

BEMT  describes the conductivity as follows:
\begin{equation}\label{eq:EffectiveMediumTheory}
f_m \frac{{\sigma _m  - \sigma_{\mathrm{eff}} }} {{\sigma _m +
\bigl( {d - 1} \bigr)\sigma_{\mathrm{eff}} }} + \bigl( {1 - f_m
} \bigr)\frac{{\sigma _i - \sigma_{\mathrm{eff}} }} {{\sigma _i +
\bigl( {d - 1} \bigr)\sigma_{\mathrm{eff}} }} = 0
\end{equation}

\noindent where \mathenv{\sigma_m} is the conductivity in the
metallic phase (\ohmcm{1000}), and \mathenv{\sigma_i=0} is the
conductivity in the insulating phase.  As with previous descriptions
of VO$_{2}$ using BEMT, we take the two-dimensional form
(\emph{d}=2) of this expression (\onlinecite {ChoiMidIRVO2}).  In
this simple model, there exist disconnected metallic domains in the
the insulating phase. Percolation of the metallic domains occurs at
\mathenv{f_m} = 0.50, at which point the sample becomes conducting.

\begin{figure}
        \includegraphics[scale=0.35,angle=0]{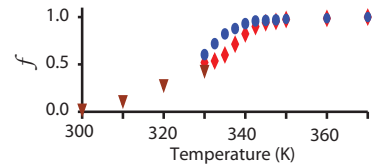}
        \caption{\label{fig:thresh}(color online)
        The volume fraction, $f_m$, responsible for the observed
        conductivity shown in \figref{fig:Sigma}
        ($\blacklozenge$ for $T_{\uparrow}$ and $\bullet$ for $T_{\downarrow}$)
        and the extracted volume fraction in the insulating state
        ($\blacktriangledown$) as determined from the dynamic
        BEMT described in the text.}
\end{figure}

We can calculate the metallic volume fraction, \mathenv{f_m} using
Eq. \eqref{eq:EffectiveMediumTheory} and the experimental results
presented in \figref{fig:Sigma}(c) for temperatures above $T_{c}$
($\blacklozenge$ for \mathenv{T_{\uparrow}} and $\bullet$ for
\mathenv{T_{\downarrow}}), which we plot in \figref{fig:thresh}. The
increasing temperature branch of the metallic fraction,
\mathenv{f_m\bigl(T_{\uparrow}\bigr)}, increases from \mathenv{0.52}
at \kelvin{330} to \mathenv{0.98} at \kelvin{350}, while the
decreasing temperature branch,
\mathenv{f_m\bigl(T_{\downarrow}\bigr)}, returns to \mathenv{0.61}
at \kelvin{330}, a consequence of the conductivity hysteresis
exhibited in this material. In the insulating phase, we cannot use
this approach to determine \mathenv{f_m} since the conductivity is
below our detection limit.

To describe the conductivity dynamics in the insulating phase using
BEMT we determine the temporal dependence of the volume fraction
using the following expression:
\begin{equation}\label{eq:dynVolFracOne}
\frac{df_m}{dt}= f_m(1-f_m)\beta(T)
\end{equation}

\noindent With this simple model, the growth rate of $f_m$ depends
on directly on $f_m$, the available nonmetallic fraction (1-$f_m$), and
$\beta(T)$ which describes the rate at which $f_m$ evolves. It is
reasonable to assume $\beta(T) = \beta_{0}\exp(-\Theta/k_{b}T)$
which describes an Arrhenius-like temperature dependence where
$\Theta$ is an energy barrier related, in the present case, to the
latent heat. For example, for homogeneous domain growth, $\Theta
\varpropto (T-T_c)^{-2}$ \cite{Sethna}. The temperature dependence
of $\beta(T)$ is important to consider as there is a bath (i.e. the
substrate) to which deposited heat can escape. This heat escape,
described as $T = T_{0}\exp(-t/\tau_{sub})$ (where tau derives from
the thermal mismatch between sample and substrate) imparts a time
dependence to $\beta(T)$. This allows for the parameterization of
$\beta(T)$ in terms of t. A subsequent Taylor expansion of
$\beta(T(t))$ about t = 0 yields an analytical solution to Eqn. 2
given by $f_m(t) = \Phi / (1+ \Phi)$ where $\Phi(t) =
f_{m}^{i}/(1-f_{m}^{i})\zeta$ where $f_{m}^{i}$ is the initial
metallic volume fraction, and $\zeta = \exp(\tau_{sub}^{r}
\beta_{0}^{r}(1-exp(-t/\tau_{sub}^{r}))$. It is the term $\zeta$ which,
even in the presence of superheating well above $T_c$, prevents the
full conductivity from being obtained. The superscript r indicates
that dimensionless factors ($>1$) from the Taylor expansion have been
incorporated into the effective lifetime and and rate to simplify
the expressions in this phenomenological model.

This solution describes the situation where the rate of increase of
$f_{m}$ decreases as energy initially deposited in the film escapes
to the substrate. This determines the rise time of the conductivity
and maximum induced change which in turn depends on the initial
temperature (immediately after heating) and the initial volume
fraction. We emphasize that this solution describes the rise time
and must be multiplied by another exponential $\exp(-t/\tau_d)$ to
include the subsequent conductivity decay (parameterized by
$\tau_d$). The solid black lines in \figref{fig:PumpProbe}(a) are
fits using this description where, for two dimensions $\sigma(t) =
(2f_{m}(t)-1)\sigma_{m}$. For the fits, $\tau_{sub}^{r}$ = 100 ps,
$\beta_{0}^{r}$ = 3.2$\times 10^{10}s^{-1}$, $\tau_d$ = 1 ns, and
$f_{m}^{i}$ = 0.08 (\kelvin{300}), $f_{m}^{i}$ = 0.13 (310K), and
$f_{m}^{i}$ = 0.3 (320K). The values of $f_{m}^{i}$ in the
insulating phase are plotted in \figref{fig:thresh}.

We see that BEMT, appropriately extended to describe a dynamic
metallic volume fraction, can account for the observed conductivity
dynamics and strongly suggests a scenario where metallic precursors
grow and coalesce upon photoinduced superheating. Furthermore, the
results display an initial condition sensitivity described by the
initial volume fraction $f_{m}^{i}$. We note that our analysis has
assumed homogeneous growth of $f_{m}$ from an initial $f_{m}^{i}$.
It is possible that there is also photoinduced nucleation in which
case the values of $f_{m}^{i}$ will be smaller than what we have
estimated from our analysis. Nonetheless, even in the case of
photoinduced nucleation, the experimental data still reveal an
initial condition sensitivity consistent with softening of the
insulating phase and the BEMT describes the essence of the observed
conductivity response.

%%%Finally, we note that it should be possible to gain additional insight
%%%into the nature of the photoinduced phase transition using spatially
%%%and temporally resolved techniques
%%%\cite{UltrafastDynamicalProcessesSemiconductors}. This would provide
%%%more direct insight into the local nature of the dynamics and enable
%%%correlation with the grain size of the films.

In summary, we studied the near-threshold behavior of the
photo\-induced phase transition in $\mathrm{VO_2}$. For the first
time, we use optical pump THz-probe measurements to directly measure
the change in conductivity of the system. The observed dynamics of
the photoinduced phase transition result in the enhancement of
fluctuations as the temperature is increased towards the transition
temperature. These results may also be conducive to high-sensitivity
optical devices, which make use of correlated oxides for switching,
detection or optical limiters.

We thank G.~T.~Wang and D.~A.~Yarotski for assistance with the film
thickness measurements and M.~Croft for an insightful discussion.
This research has been supported by the Los Alamos National
Laboratory Directed Research and Development program.

\end{document}